\documentclass[12pt,a4paper,dvipdfmx]{article}
\usepackage{graphicx}
\usepackage{amssymb,amsfonts,amsmath}
\usepackage{cite}
\usepackage{bm}

\usepackage{array}
\newcolumntype{x}[1]{>{\centering\arraybackslash\hspace{0pt}}p{#1}}

\title{Opinion control in complex networks}
\bigskip
\author{Naoki Masuda${}^{1*}$
\ \\
\ \\
${}^{1}$ 
Department of Engineering Mathematics, University of Bristol,\\
Merchant Venturers Building,\\
Woodland Road, Clifton, Bristol BS8 1UB, United Kingdom
\ \\
$^*$ Corresponding author (naoki.masuda@bristol.ac.uk)}

\begin{document}
\setlength{\baselineskip}{24pt}

\maketitle

\newpage

\section*{Abstract}
In many instances of election, the electorate appears to be a composite of partisan and independent voters. Given that partisans are not likely to convert to a different party, a main goal for a party could be to mobilize independent voters toward the party with the help of strong leadership, mass media, partisans, and effects of peer-to-peer influence. Based on the exact solution of the classical voter model dynamics in the presence of perfectly partisan voters (i.e., zealots), we propose a computational method to maximize the share of the party in a social network of independent voters by pinning control strategy. The party, corresponding to the controller or zealots, optimizes the nodes to be controlled given the information about the connectivity of independent voters and the set of nodes that the opponent party controls. We show that controlling hubs is generally a good strategy, whereas the optimized strategy is even better. The superiority of the optimized strategy is particularly eminent when the independent voters are connected as directed rather than undirected networks.

\newpage

\section{Introduction}

We often flip our opinions in response to what others are doing. In various situations ranging from voting to adoption of new habits, it is widely recognized that opinion formation on a large scale is considerably influenced by
peer-to-peer interaction among individuals embedded in social networks \cite{Barrat2008book,Jackson2008book,Castellano2009RevModPhys,Easley2010book}. On the other hand, such contagion effects seem to be irrelevant to some individuals that are resolute on the matter and do not feel the pressure of peers.
For example, in election, the electorate is usually a composition of partisan voters in favor of specific parties and independent voters \cite{Keith1992book,Bartels2000AmJPolitSci}. Because partisans are reluctant to change, the main goal for a party would be to mobilize independent voters toward its candidate. The aim in the present theoretical study is to explore efficient strategies for attracting independent voters when independent voters are connected as a social network and different parties are allowed to influence some independent voters.

In models of collective opinion formation, perfectly partisan voters, or the parties themselves, have been analyzed under the names of pinning controllers and zealots.
Pinning control of networks, in which a controller unidirectionally affects some selected nodes, is an effective method to guide
the state of nodes in a coupled dynamical system to a desired state such as the synchronized state. The idea has been explored in linear and nonlinear dynamical systems coupled in networks \cite{Sorrentino2007PhysRevE,Porfiri2008Automatica,Delellis2010IEEEMag}.
We apply this idea to opinion control in complex networks. Theoretical studies of pinning opinion control in collective opinion dynamics date back to at least studies of zealots in the voter model. 
The voter model is a paradigmatic model of collective opinion formation, in which individuals would flip their opinions at a rate proportional to the number of neighboring others selecting the opposite opinion\cite{Liggett1985book,Redner2001book,Barrat2008book,Castellano2009RevModPhys,Krapivsky2010book}. Mobilia first showed that dynamic dramatically changed upon the introduction of zealots, i.e., stubborn voters, to the population of ordinary voters \cite{Mobilia2003PhysRevLett}.
Zealots correspond to perfect partisans, leaders in a community, or mass media and are mathematically the same as pinning controllers (see \cite{MasudaGibertRedner2010PhysRevE,MasudaRedner2011JStatMech} for the case of error-prone zealots). With the voter model and other opinion formation models, it has been also shown that existence of zealots in favor of the opposite opinions, like competing parties,
induces coexistence of the different opinions in the equilibrium
\cite{WuHuberman2004arxiv,Mobilia2005PhysRevE,Chinellato2007arxiv,Mobilia2007JStatMech,AcemogluComo2011IEEEConf,Yildiz2013ACMEconComput}. This phenomenon contrasts to that for the voter model without zealots \cite{Liggett1985book,Barrat2008book,Castellano2009RevModPhys} and with a single type of zealot
\cite{Mobilia2003PhysRevLett}; in these cases consensus is always reached in finite networks and low-dimensional infinite lattices.

Controlling opinion dynamics implies that there is a control objective such as the desired final state or cost minimization.
Most of existing literature on voter models in the presence of zealots has described
phenomena induced by zealots but has not treated opinion control in a proactive sense.
Among the studies that treated opinion control explicitly, some studies
assumed the presence of just a single type of zealot
\cite{QianCao2011Chaos,Zuev2012AutomRemCtrl,Hegselmann2014SSRN,Eymann2014preprint}.
To the best of our knowledge, two studies considered opinion control in networks by competing zealots, similarly to the present study.
In Ref.~\cite{Kuhlman2013CompNetw}, the authors considered a control strategy in which the highest-degree nodes
are controlled. They solved an optimization problem numerically by transforming the problem into that of a random walk in the network and numerically running the random walk. Their objective function was the control cost under the condition that the fraction of votes in favor of the controller's opinion must be at least a prescribed threshold.
In Ref.~\cite{Yildiz2013ACMEconComput}, the authors treated optimal selection of newly controlled nodes under the condition that the two subsets of nodes that the focal zealot and the opposing type of zealot controlled at the moment were known.
They proposed a greedy algorithm that exploited the submodularity of the objective function. However, the main focus of the article was in theoretical evaluation of the algorithm, and the proposed algorithm was not numerically tested on networks.

In contrast to these studies \cite{Kuhlman2013CompNetw,Yildiz2013ACMEconComput}, in the present study we establish the procedure to maximize the number of vote by independent voters using linear algebra. We assume zealots (equivalently, pinning controllers) possessing different opinions can influence some independent voters, represented by nodes in a social network, to try to coax independent voters into their own parties. On the basis of the exact solution on the mean number of vote in favor of each opinion in the presence of zealots, we establish a method to compute the solution and heuristically optimize the mean number of vote using power iteration. Then, we apply the proposed method to several complex networks. In particular, we examine whether controlling high-degree nodes (i.e., hubs) is efficient or not depending on networks.

\section{Model}

We assume a directed and weighted network with $N$ nodes, which may have self loops. Each node is occupied by an independent voter. Therefore, we also refer to nodes as independent voters.
The weight of the link from node $i$ to $j$ is denoted by $w_{ij}\ge 0$ and represents the strength of the influence of independent voter $i$ on independent voter $j$ in opinion formation dynamics.
The $N$ independent voters dynamically switch their opinions between two options A and B, as we will specify in the following.

We also assume that
the network is strongly connected so that there is no root node or root component to send out directed links to the remainder of the network without receiving directed links from outside. Such a root would effectively function as zealot because the opinion of the root is not affected by the remainder of the network; we will introduce zealots externally in the following. For undirected networks, the strong connectedness is reduced to the usual connectedness.

\begin{figure}[h]
\begin{center}
\includegraphics[width=12cm]{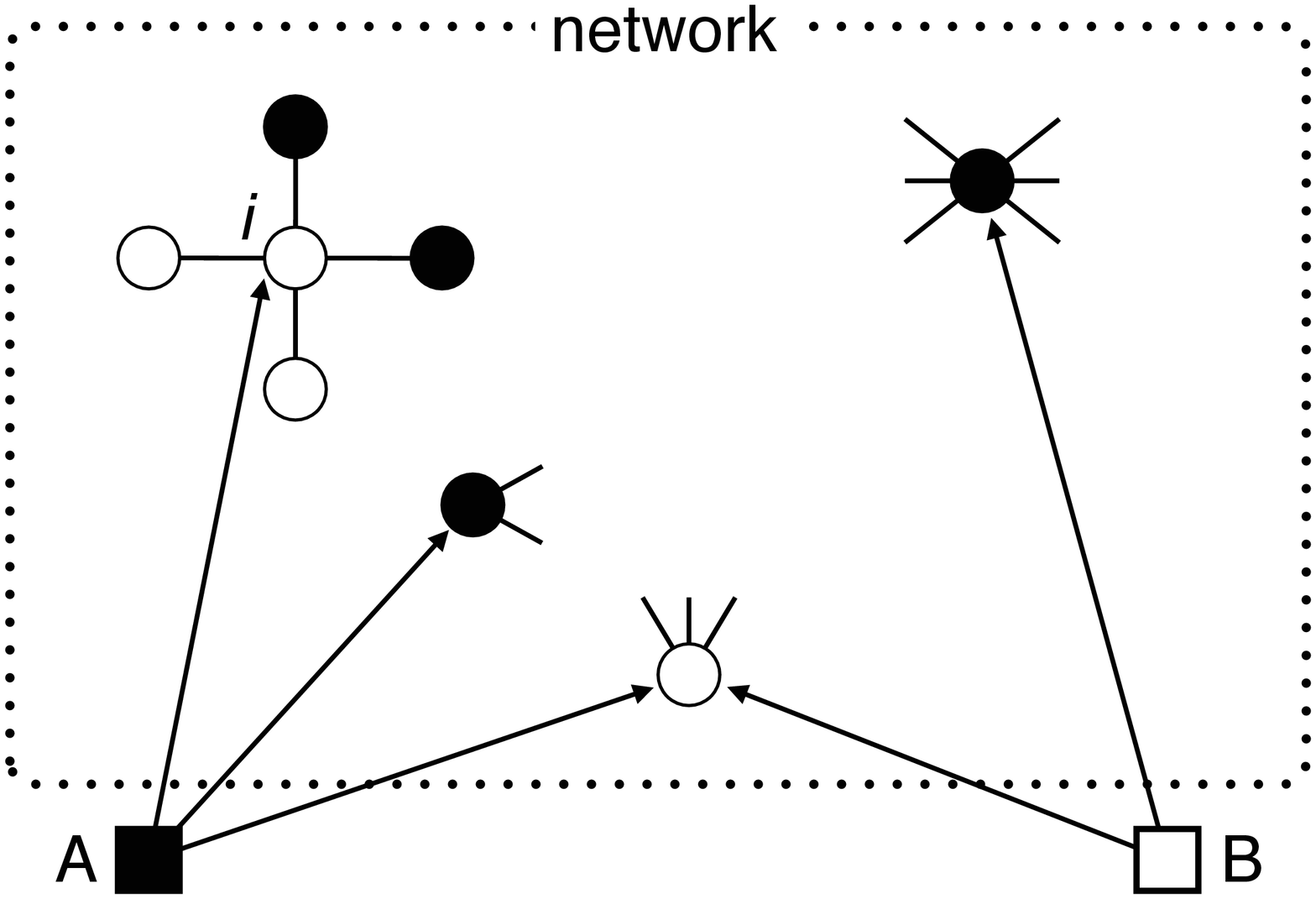}
\caption{Schematic of opinion control and opinion updating. The circles represent independent voters. The squares represent zealots.
The filled and open symbols correspond to opinions A and B, respectively. Upon updating, node $i$ adopts opinions A and B with probability $3/5$ and $2/5$, respectively.}
\label{fig:schem}
\end{center}
\end{figure}

Apart from the $N$ independent voters, we assume zealots (i.e., perfect partisans) each of which favors one opinion.
By definition, zealots never change the opinion and are given as elements external to the network of independent voters (Fig.~\ref{fig:schem}). The control gain represents the strength with which a zealot influences an independent voter. We denote by $p_{A,i} (\ge 0)$ and $p_{B,i} (\ge 0)$ the control gain on independent voter $i$ for a zealot in favor of opinion A and B, respectively. The control gain plays the same role as the link weight, $w_{ij}$. We use different terms and symbols to distinguish interaction between a pair of independent voters (link weight, $w_{ij}$) and that between a zealot and an independent voter (control gain, $p_{A,i}$ and $p_{B,i}$).
We consider a single zealot in favor of opinion A (called A zealot in the following) and that in favor of B (B zealot) such that they influence some independent voters, perhaps with different control gains. We do not lose generality by assuming a single zealot of each type because two A zealots influencing independent voter $i$ with a unity control gain is equivalent to one A zealot influencing $i$ with a control gain equal to two.

From zealots' perspectives, each ``party'' (i.e., zealot) may be interested in wisely selecting $p_{A,i}$ and $p_{B,i}$ ($1\le i\le N$) under given constraints. For example, the sum of the control gain may be upper-bounded because the cost of control is proportional to the total gain. Alternatively, each type of zealot may be able to control a given number of nodes each with a unity gain if the party has to make a binary decision regarding whether or not to persuade each independent voter and accessing individual independent voters is a costly process.

In each update event, one of the independent voters in the network, node $i$, is selected with equal probability $1/N$. Then, $i$ copies the opinion of one of its upstream neighbors in the network (usual neighbors in the case of undirected network) or that of a zealot if $i$ is directly influenced by it. The independent voter or the zealot whose opinion $i$ mimicks is selected with the probability proportional to the link weight or the control gain. Figure~\ref{fig:schem} is a schematic example of an undirected and unweighted network. Suppose that node $i$, which has degree four and is controlled by the A zealot (with the unity gain assumed, i.e., $p_{A,i}=1$), is selected for updating. Among the four neighboring independent voters and the A zealot, three possess opinion A (filled circles and square) and two opinion B (open circles). Therefore, $i$ adopts opinion A and B with probability $3/5$ and $2/5$, respectively.
Generally speaking, a selected node $i$ selects the opinion of independent voter $j$ for copying with probability $w_{ji}/\left(\sum_{\ell=1}^N w_{\ell i} + p_{A,i} + p_{B,i}\right)$, the A zealot with probability $p_{A,i}/\left(\sum_{\ell=1}^N w_{\ell i} + p_{A,i} + p_{B,i}\right)$, and the B zealot with probability $p_{B,i}/\left(\sum_{\ell=1}^N w_{\ell i} + p_{A,i} + p_{B,i}\right)$. The update rule is the same as the so-called VM (voter model) rule among the three main update rules proposed in Refs.~\cite{Antal2006PhysRevLett,Sood2008PhysRevE}. The other update rules are briefly discussed in section~\ref{sec:discussion}.

Each node is updated once per unit time on average. Owing to the presence of zealots in the opposite opinions, the consensus (also called fixation) of either opinion never occurs \cite{WuHuberman2004arxiv,Mobilia2005PhysRevE,Chinellato2007arxiv,Mobilia2007JStatMech,AcemogluComo2011IEEEConf,Yildiz2013ACMEconComput}. We are interested in the fraction of nodes in favor of each opinion in the equilibrium.

\section{Results}

\subsection{Meanfield analysis}\label{sec:meanfield}

We start with a meanfield analysis assuming a well-mixed infinite population corresponding to an undirected and unweighted network of infinite size. For this case, more elaborate theoretical results regarding distributions of
the number of independent voters in either opinion are available \cite{WuHuberman2004arxiv,Chinellato2007arxiv,Mobilia2007JStatMech}.
Denote by $x$ ($0\le x\le 1$) the fraction of independent voters possessing opinion A. The fraction of independent voters possessing opinion B is equal to $1-x$. We assume that there are an additional fraction of $a$ and $b$ zealots
in favor of A and B, respectively, where the fraction refers to that relative to the number of independent voters. For example, $a=0.1$ implies that the number of A zealot is equal to 10\% of the number of the independent voters. We also assume that the zealots affect each independent voter with a unity gain.
Finally, we assume $a, b > 0$.
Although the original model has a single zealot in each opinion, in this section we translated it to the fraction of zealots because
a single zealot does not have an influence in an infinite population unless the number of independent voters that the zealot influences or the control gain is infinite.

In this situation, the opinion formation dynamic is given by
\begin{equation}
\frac{{\rm d}x}{{\rm d}t} = (1-x)(x + a) - x (1-x + b),
\label{eq:master meanfield}
\end{equation}
where $t$ is time.
The steady state is given by
\begin{equation}
x^* = \frac{a}{a+b}.
\label{eq:x^*}
\end{equation}
This steady state is stable with the eigenvalue $-a-b < 0$; if the fraction of zealot is large, the convergence to the equilibrium given by Eq.~\eqref{eq:x^*} is fast.
Equation~\eqref{eq:x^*} implies that the fraction of the opinion of independent voters in the equilibrium is equal to that of zealots.

\subsection{Average opinion of independent voters in general networks}

Denote by $u_{A,i}$ ($1\le i\le N$) the probability that independent voter $i$ takes opinion $A$. The probability that $i$ takes opinion $B$ is given by $u_{B,i} = 1 - u_{A,i}$. The master equation of the opinion dynamics is given by
\begin{equation}
\frac{{\rm d}u_{A,i}}{{\rm d}t} =
(1-u_{A,i}) \frac{\sum_{j=1}^N w_{ji} u_{A,j} + p_{A,i}}
{\sum_{\ell=1}^N w_{\ell i} + p_{A,i} + p_{B,i}}
- u_{A,i} \frac{\sum_{j=1}^N w_{ji} (1-u_{A,j}) + p_{B,i}}
{\sum_{\ell=1}^N w_{\ell i} + p_{A,i} + p_{B,i}}.
\label{eq:master equation for u_{A,i}}
\end{equation}
In the steady state, we set the left-hand side of Eq.~\eqref{eq:master equation for u_{A,i}} to zero to obtain
\begin{equation}
\left[L + {\rm diag}\left(p_{A,1}+p_{B,1}, \ldots, p_{A,N}+p_{B,N}\right)\right]
\begin{pmatrix}u_{A,1}\\ \vdots\\ u_{A,N}\end{pmatrix}
=
\begin{pmatrix}p_{A,1}\\ \vdots\\ p_{A,N}\end{pmatrix}.
\label{eq:u_{A,i}}
\end{equation}
Matrix $L=(L_{ij})$ is the $N\times N$ Laplacian of the network
defined by
\begin{equation}
L_{ij} = \delta_{ij}\sum_{\ell=1}^N w_{\ell i} - (1-\delta_{ij}) w_{ji}\quad (1\le i, j\le N),
\label{eq:def L}
\end{equation}
where $\delta_{ij}$ is the kronecker delta. In Eq.~\eqref{eq:u_{A,i}},
${\rm diag}$ represents the $N\times N$ diagonal matrix whose diagonal elements are specified by the arguments.
In fact, $u_{A,i}$ is equivalent to the $i$th element of $\overline{\bm u}_1^3$ in
Ref.~\cite{MasudaKori2010PhysRevE} derived from the Laplacian matrix for the extended network including the zealots, where we interpret $m=2$, $b=3$, and the A zealot, the B zealot, and the strongly connected network of independent voters correspond to the first, second, and third blocks, respectively, of the irreducible normal form given by Eq.~(9) in Ref.~\cite{MasudaKori2010PhysRevE}. It should also be noted that Eq.~\eqref{eq:u_{A,i}} above is equivalent to Eq.~(3) in Ref.~\cite{AcemogluComo2011IEEEConf}, which the authors derived by assuming that every independent voter was controlled by a zealot (i.e., $p_{A,i}+p_{B,i}>0$ for all $i$).

The share of opinion A in the steady state, denoted by $S_A$, is given by
\begin{equation}
S_A = \frac{\sum_{i=1}^N u_{A,i}}{N}.
\label{eq:influence of zealot}
\end{equation}
The steady state, $u_{B,i}$ ($1\le i\le N$), and the share of opinion B in the steady state, $S_B$, are given by
Eqs.~\eqref{eq:u_{A,i}} and \eqref{eq:influence of zealot}, respectively, with A replaced by B.
It should be noted that $u_{A,i}+u_{B,i} = 1$ ($1\le i\le N$) and hence $S_A + S_B = 1$ (Appendix~A). 

Equation~\eqref{eq:master equation for u_{A,i}} can be rewritten as
\begin{equation}
\frac{{\rm d}u_{A,i}}{{\rm d}t} =
\frac{\sum_{j=1}^N w_{ji} (u_{A,j} - u_{A,i}) + p_{A,i}(1-u_{A,i})  + p{B,i}(0-u_{B,i})}
{\sum_{\ell=1}^N w_{\ell i} + p_{A,i} + p_{B,i}}.
\label{eq:DeGroot}
\end{equation}
This is the continuous-time DeGroot model \cite{Olfatisaber2007IEEE,MasudaKawamuraKori2009NewJPhys}, if we interpret $u_{A,i}$ as the continuously valued opinion, with one and zero corresponding to the pure opinion A and pure opinion B, respectively. It should be noted that Eq.~\eqref{eq:DeGroot} implies that the independent voters attract each other and are also attracted to A (i.e., 1) and B (i.e., 0) with gain $p_{A,i}$ and $p_{B,i}$, respectively. 

\subsection{Toy examples}

Consider the undirected and unweighted complete graph with control gains $p_{A,i} = a$ and $p_{B,i} = b$ ($1\le i\le N$).
We substitute $L_{ij} = \delta_{ij}(N-1)
- (1-\delta_{ij})$ in Eq.~\eqref{eq:u_{A,i}} to obtain
$u_{A,i}=a/(a+b)$ ($1\le i\le N$). This result is consistent with that obtained from the meanfield analysis in section~\ref{sec:meanfield}.

\begin{figure}[h]
\begin{center}
\includegraphics[width=6cm]{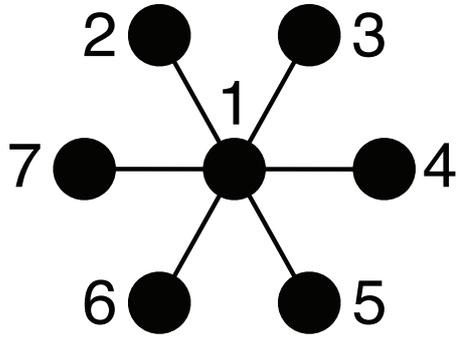}
\caption{Star graph having seven nodes.}
\label{fig:star schem}
\end{center}
\end{figure}

Consider a undirected and unweighted star with $N$ nodes (Fig.~\ref{fig:star schem}). Node 1 is the unique hub adjacent to all other nodes, each of which is only adjacent to the hub. Assume that the B zealot controls node 2 with gain $b$ (i.e., $p_{B,i}=b\delta_{i,2}$) and that the A zealot controls either the hub or a non-hub with gain $a$.
Equation~\eqref{eq:u_{A,i}} yields the following results. If the A zealot controls the hub, we obtain
$u_{A,i} = (ab+a)/(ab+a+b)$ ($i\neq 2$), $u_{A,2} = a/(ab+a+b)$, and
$S_A = \left[((N-1)ab/N + a\right]/(ab+a+b)$.
If the A zealot controls node 2, we obtain
$u_{A,i} = a/(a+b)$ ($1\le i\le N$) and $S_A = a/(a+b)$.
If the A zealot control single node $i$ ($3\le i\le N$), we obtain
$u_{A,i}= (ab+a)/(2ab+a+b)$ ($i=1, 4\le i\le N$), $u_{A,2} = a/(2ab+a+b)$, $u_{A,3}= (2ab+a)/(2ab+a+b)$,
and $S_A = (ab+a)/(2ab+a+b)$.

\begin{figure}[h]
\begin{center}
\includegraphics[width=8cm]{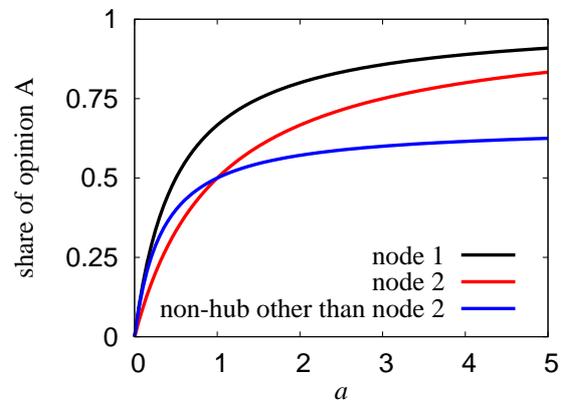}
\caption{Share of opinion A in the star graph. We set $b=1$, $N\to\infty$, and vary $a$. The legend represents the node that the A zealot controls. The B zealot controls node 2 in all cases.}
\label{fig:star results}
\end{center}
\end{figure}

The behavior of the share of opinion A (i.e., $S_A$) when we set $b=1$, $N\to\infty$, and vary $a$ is shown in
Fig.~\ref{fig:star results}. As expected, $S_A$ increases with $a$ in all cases.
When the A zealot controls the hub (black line), $S_A$ is larger than when the A zealot controls
a non-hub (red and blue lines). When the A zealot controls a non-hub node, $S_A$ is equal to 0.5 at $a=1$ because the other zealot also controls a non-hub with the same gain. However, when $a\neq 1$, $S_A$ depends on whether the non-hub nodes that the two zealots control are the the same (red line) or different (blue line).

\subsection{Numerical results}

In this section, we numerically evaluate the effect of control protocols on one model network and three real-world networks. We do not assume self connection in the following.

\subsubsection{Numerical procedure}

We numerically calculate $S_A$ and $S_B$ as follows.
Because $L + {\rm diag}\left(p_{A,1}+p_{B,1}, \ldots, p_{A,N}+p_{B,N}\right)$ is diagonally dominant, we can obtain its solution by iteration methods such as the Jacobi and Gauss-Seidel iteration \cite{Golub1996book}.
Starting from some initial conditions, which we set to $u_{A,i}=u_{B,i}=1/2$ ($1\le i\le N$), we repeat
\begin{equation}
u_{A,i} \gets \frac{p_{A,i} + \sum_{j=1}^N w_{ji}u_{A,j}}
{\sum_{\ell=1}^N w_{\ell i} + p_{A,i} + p_{B,i}},
\label{eq:iteration scheme}
\end{equation}
where $\gets$ means that we substitute the right-hand side in the left-hand side. 
We carry out Eq.~\eqref{eq:iteration scheme} simultaneously for all $i$
in the case of the Jacobi iteration, which we use in the following numerical simulations. Finally,
we use Eq.~\eqref{eq:influence of zealot} to calculate $S_A$ (and $S_B = 1 - S_A$).

\subsubsection{Barab\'{a}si-Albert (BA) model}

We start with the Barab\'{a}si-Albert (BA) model network having $N=200$ nodes. The model produces
scale-free networks with the power-law exponent of the degree distribution equal to three when $N\to\infty$ \cite{Barabasi1999Science}. To generate a network, we assume that there are initially
two nodes that are connected by a link and that any arriving node brings in two links.
The generated network has 397 undirected and unweighted links.

\begin{figure}[h]
\begin{center}
\includegraphics[width=8cm]{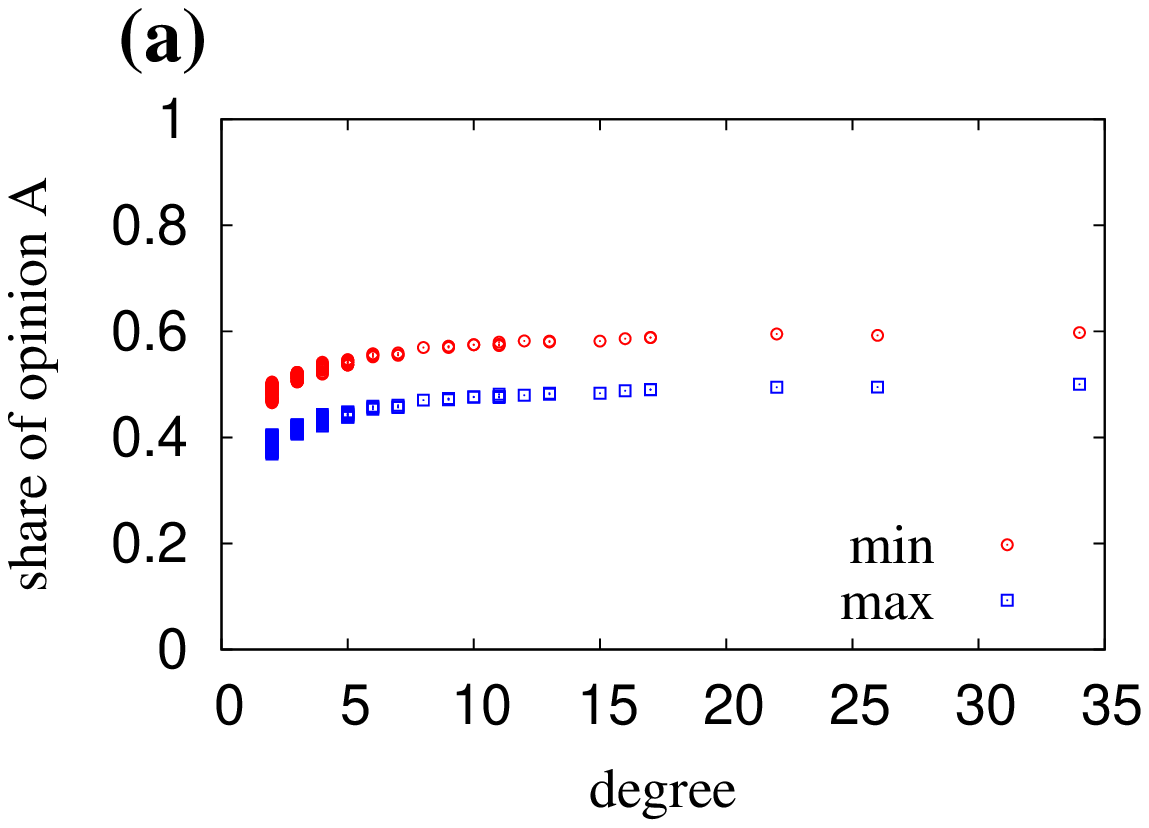}
\includegraphics[width=8cm]{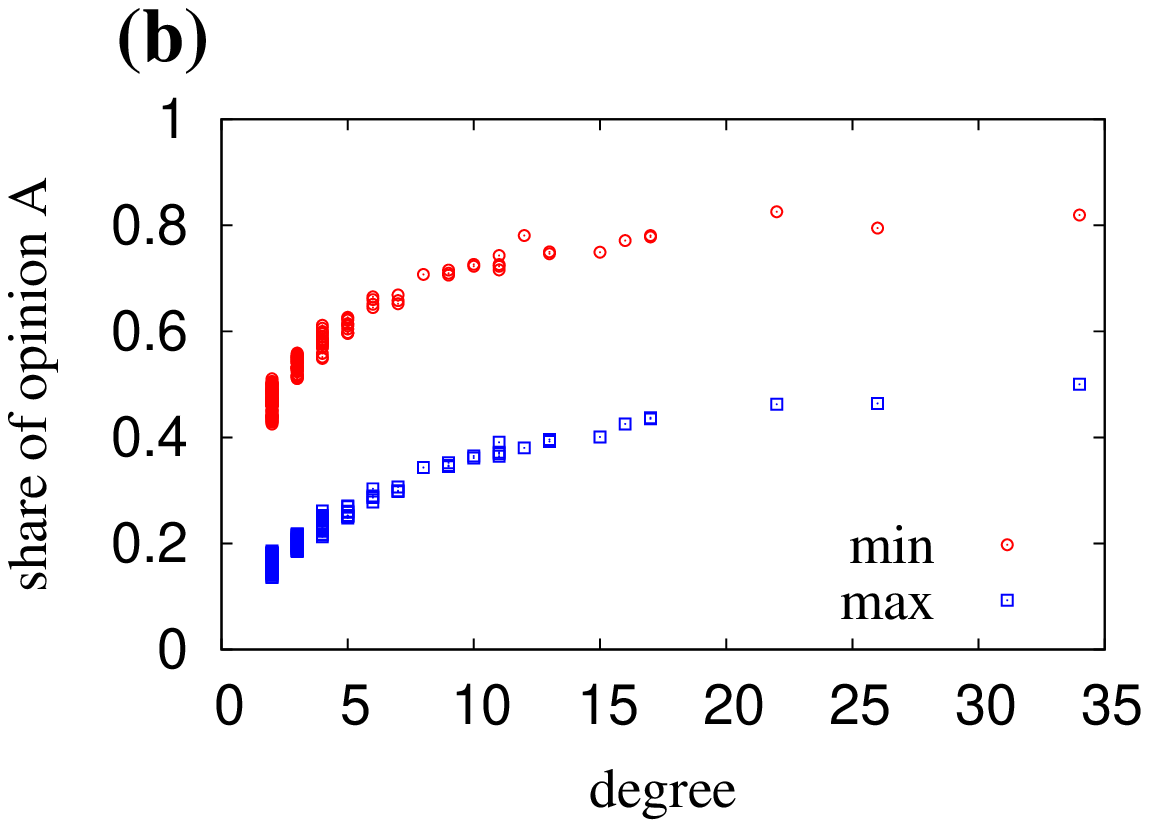}
\includegraphics[width=8cm]{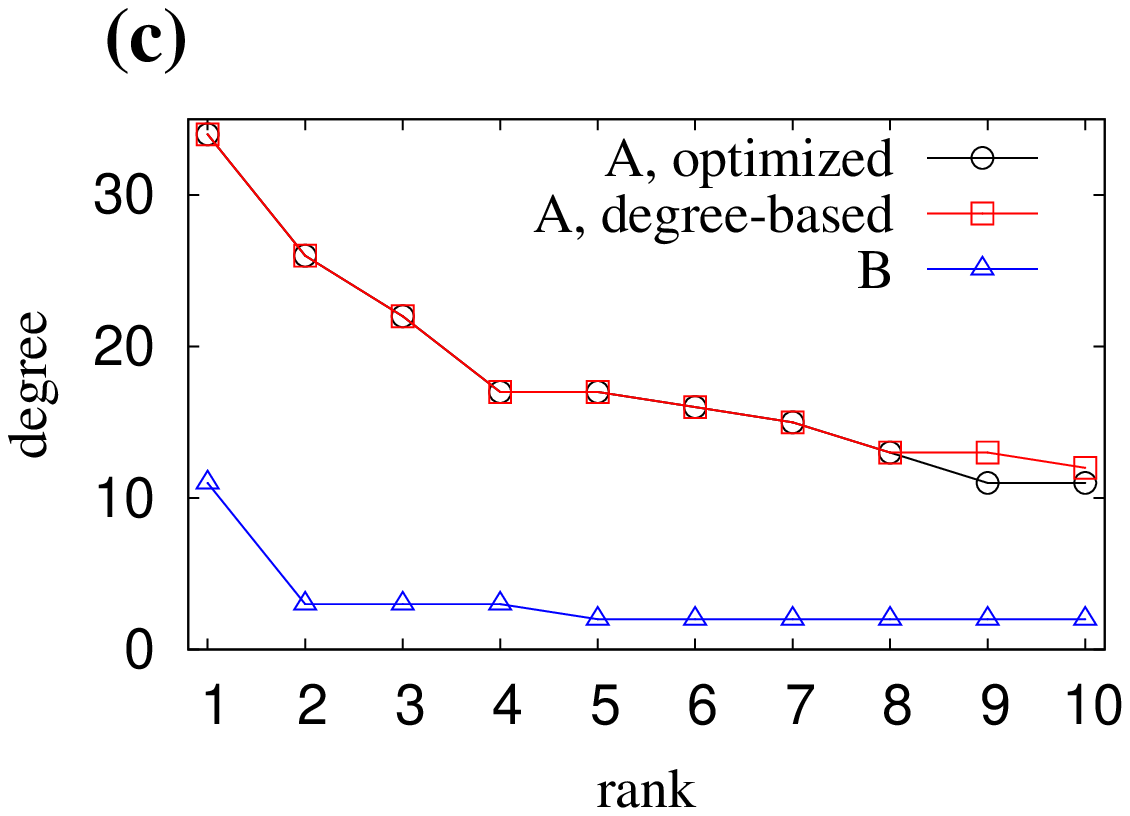}
\includegraphics[width=8cm]{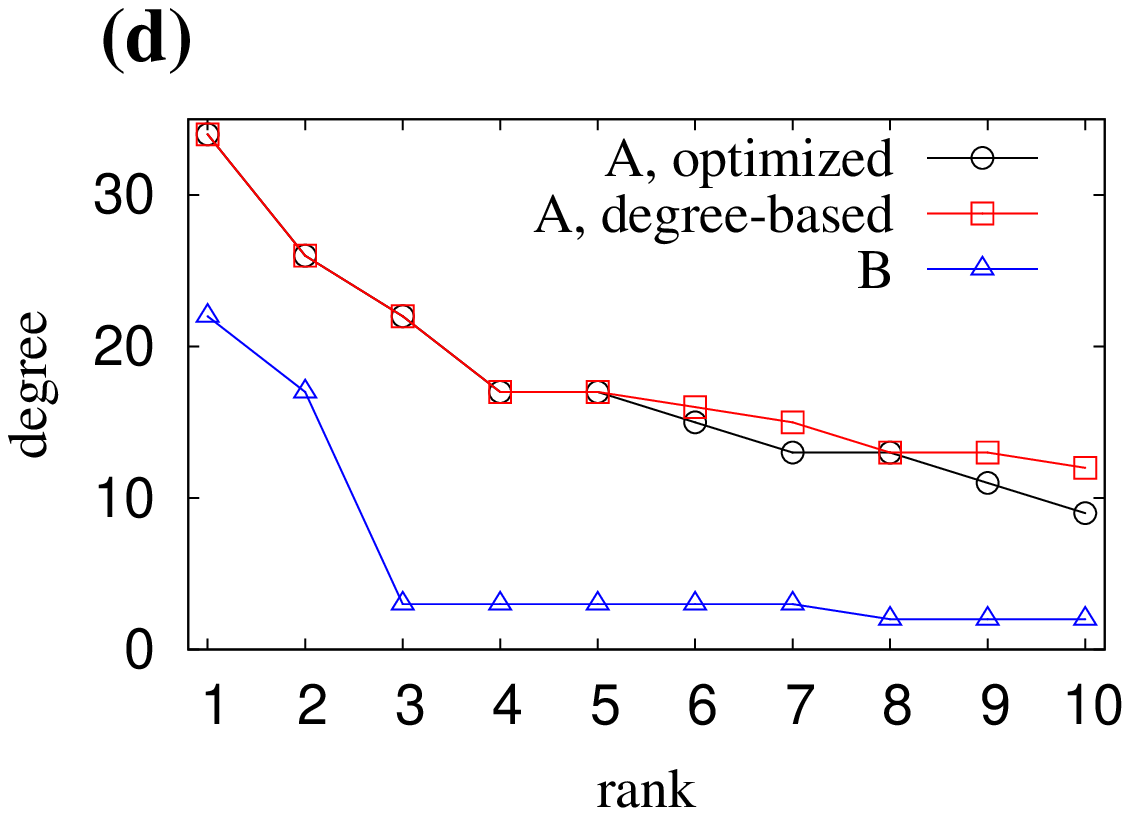}
\caption{Results for the BA model with $N=200$ nodes and 397 links.
The results when each zealot controls a single node are shown in (a) and (b). The control gain is set to 1 in (a) and 10 in (b) for both zealots. The circles and squares in these panels represent the cases in which the B zealot controls the node with the smallest and largest degree, respectively. The value on the horizontal axis represents the degree of the node that the A zealot controls. In (c) and (d),  the degrees of the controlled nodes when each zealot controls ten nodes are shown in the descending order. The rank refers to that in terms of the degree.
Given randomly selected ten nodes that the B zealot controls, whose degrees are shown by the triangles,
the optimized A zealot controls ten nodes whose degrees are shown by the circles. The degrees of the ten largest-degree nodes, which the A zealot using the degree-based protocol controls, are shown by the squares.
The control gain is set to 1 in (c) and 10 in (d) per node for both zealots. In (c) and (d), we have used different sets of randomly selected ten nodes for the B zealot. The degrees shown by the squares in (c) and (d) are the same.}
\label{fig:ba}
\end{center}
\end{figure}

First, we examine the case in which both zealots control a single node with the same gain.
The B zealot is assumed to control the node with either the largest or the smallest degree. If there are multiple nodes with the same degree, we randomly select one.
The share of opinion A when the A zealot controls one of the $N$ nodes
is shown in Figs.~\ref{fig:ba}(a) and \ref{fig:ba}(b) for the gain equal to 1 and 10, respectively.
In these figures, the circles and squares represent the results 
when the B zealot controls the node with the smallest and largest degree, respectively. 
The nodes controlled by the B zealot are the same between Figs.~\ref{fig:ba}(a) and \ref{fig:ba}(b).

Figure~\ref{fig:ba} indicates that the share of opinion A is large when the A zealot controls a node
with a large degree. The share of A is larger when the B zealot controls the node with the smallest degree (circles) than that with the largest degree (squares). These results indicate that it is efficient to control hubs rather than small-degree nodes, which is consistent with the previous results \cite{WuHuberman2004arxiv,Kuhlman2013CompNetw}.
However, the degree is a dominant but not the sole determinant of the share of A, consistent with a previously made remark
\cite{Yildiz2013ACMEconComput}.
Controlling nodes with the same degree generally leads to different $S_A$ values. In addition,
in Fig.~\ref{fig:ba}(b), $S_A$ is slightly larger for the node
with degree 22 than the largest-degree node (i.e., degree 34) when the B zealot controls the node with the smallest degree (circles). As a separate observation,
for a larger control gain (Fig.~\ref{fig:ba}(b)), the share of A is more sensitive to
the degree than for a smaller control gain (Fig.~\ref{fig:ba}(a)). This is intuitively because a large control
gain implies that the influence of the zealots is relatively important in opinion formation dynamics as compared to peer-to-peer interaction between independent voters.

Next, we assume that each zealot controls ten nodes with the same gain and examine the case in which the A zealot optimizes the set of nodes to be controlled. The B zealot is assumed to control ten randomly selected nodes.
The A zealot is allowed to use this information. First, the A zealot starts with controlling
randomly selected ten nodes. Second, we calculate the share of opinion A, $S_A$.
Third, we tentatively swap one randomly selected controlled node with one randomly selected uncontrolled node. Fourth, we recalculate $S_A$. If the new $S_A$ value is larger than the old value, then we adopt the swapping.
Otherwise, we discard the swapping. We repeat swapping attempts $2\times 10^4$ times unless otherwise stated. 

The $S_A$ values after the optimization are shown in Table~\ref{tab:optimized}
for the control gain equal to 1, 10, and 100. In the table,
we also show the results for the degree-based protocol \cite{Kuhlman2013CompNetw}, which is defined by the A zealot controlling the ten nodes with the largest degrees. We break ties by randomly selecting the required number of nodes with the threshold degree. For example, suppose that the network is undirected and unweighted, there are nine nodes whose degree is at least 20, and there are three nodes with degree 19. Then, we control one of the three nodes with degree 19 and the nine nodes with the larger degrees. 
For each gain value, the ten nodes that the B zealot controls is the same between the optimized and degree-based A zealot. Table~\ref{tab:optimized} suggests that $S_A$ after optimization is slightly larger than that obtained from the degree-based protocol. The degree of the ten controlled nodes by the optimized A zealot, degree-based A zealot, and the B zealot are shown in Figs.~\ref{fig:ba}(c) and \ref{fig:ba}(d) for the gain equal to 1 and 10, respectively. The B zealot mostly controls nodes with small degrees owing to the scale-free property of the BA model. The optimized A zealot tends to control hubs. However, some nodes that the optimized A zealot controls are not among the ten largest-degree nodes.

\subsubsection{Coauthorship network}

\begin{figure}[h]
\begin{center}
\includegraphics[width=8cm]{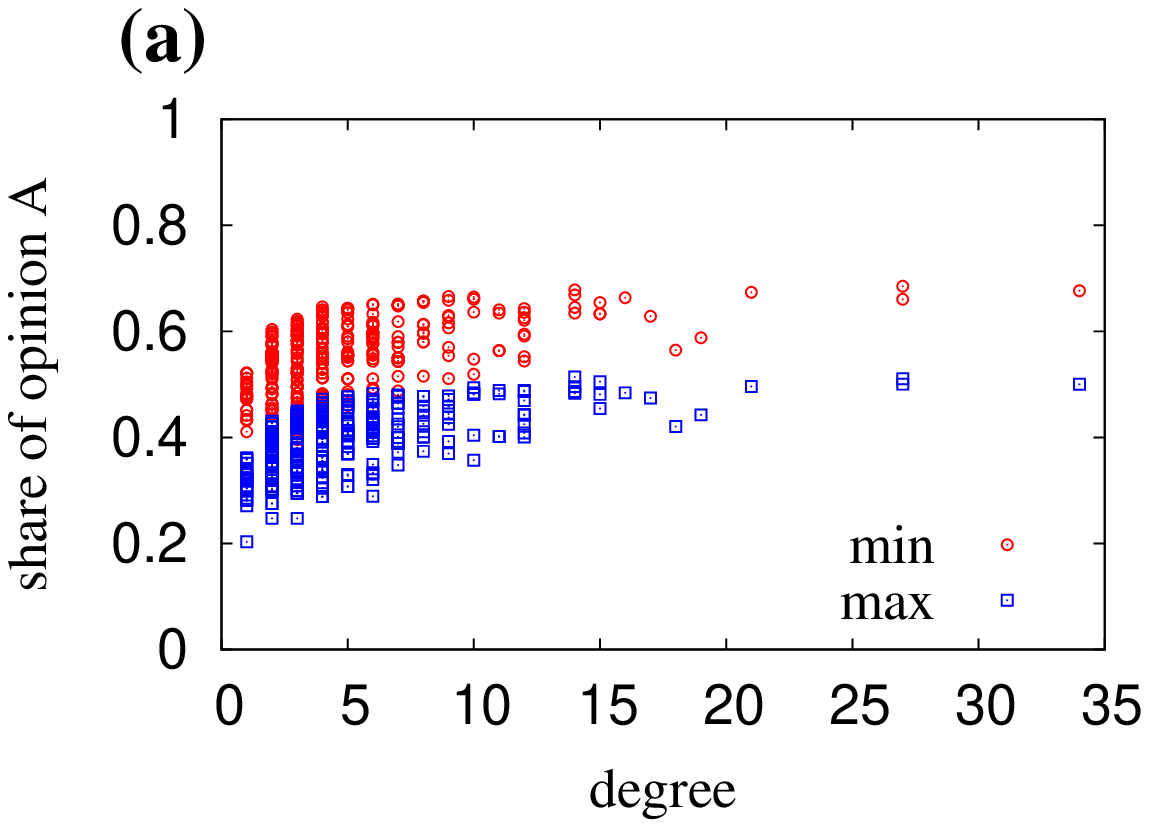}
\includegraphics[width=8cm]{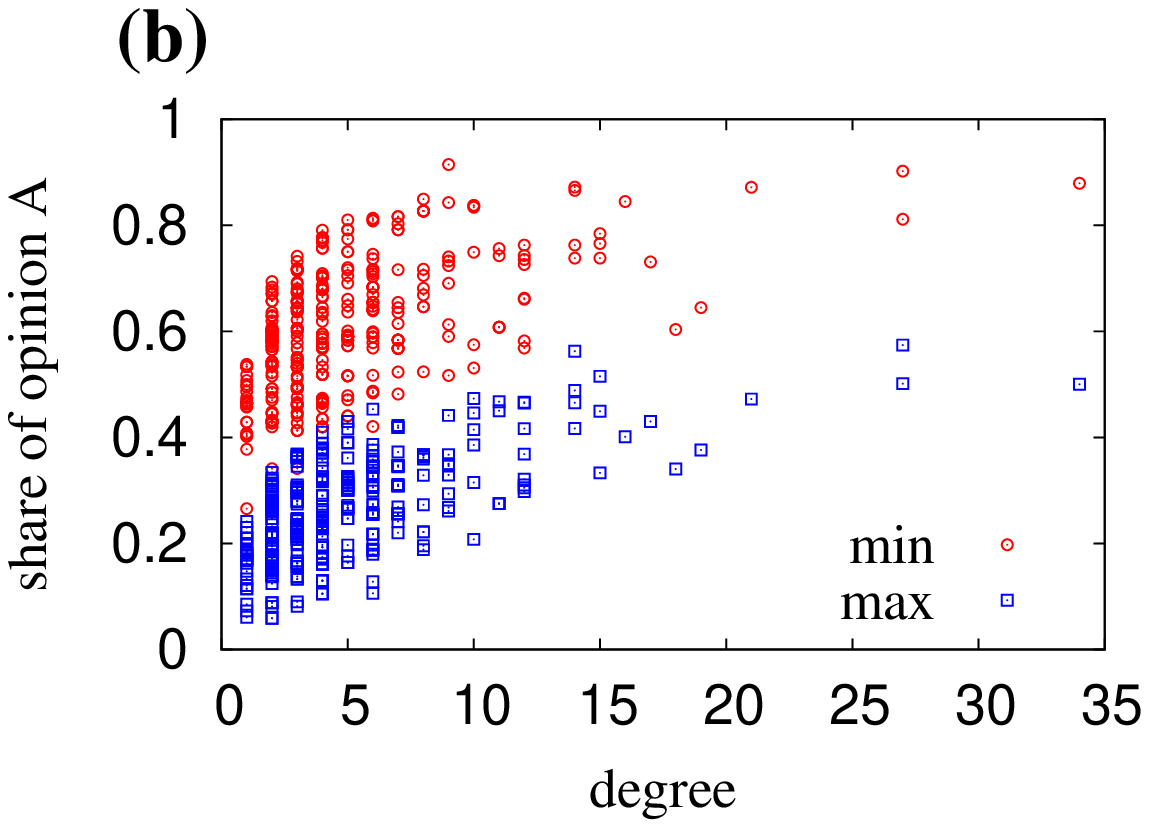}
\includegraphics[width=8cm]{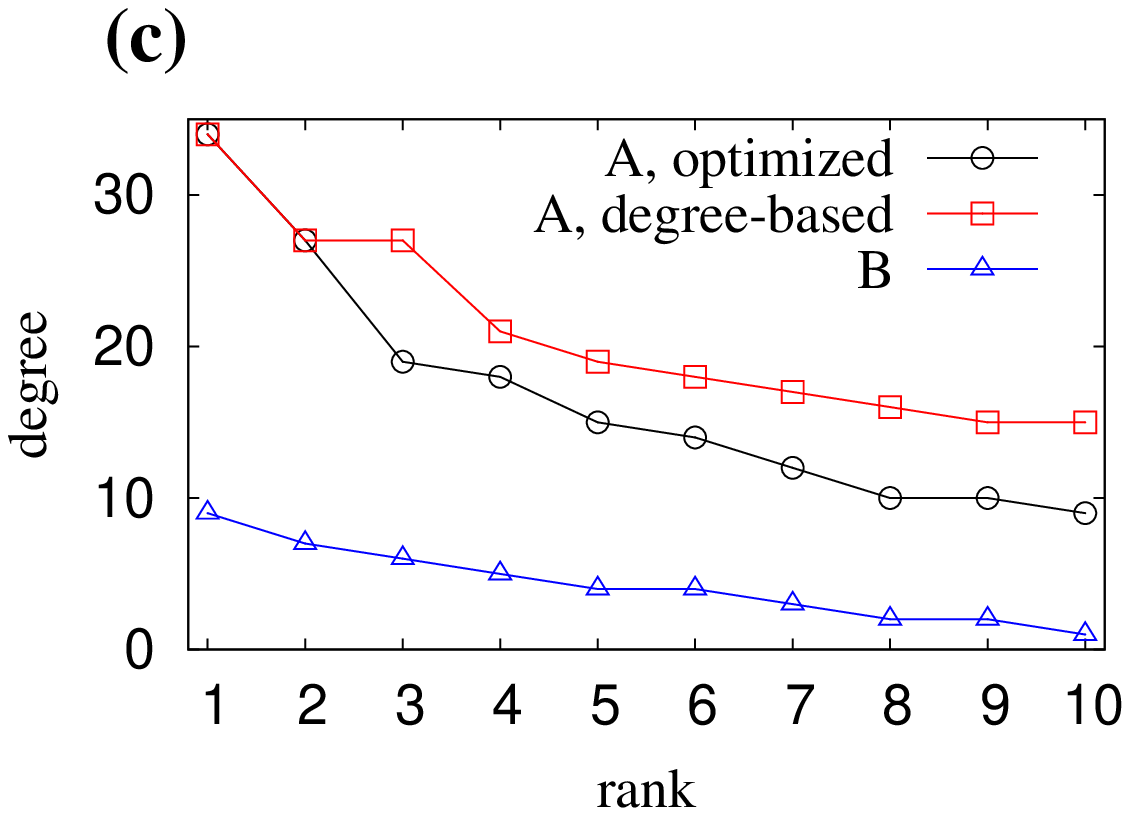}
\includegraphics[width=8cm]{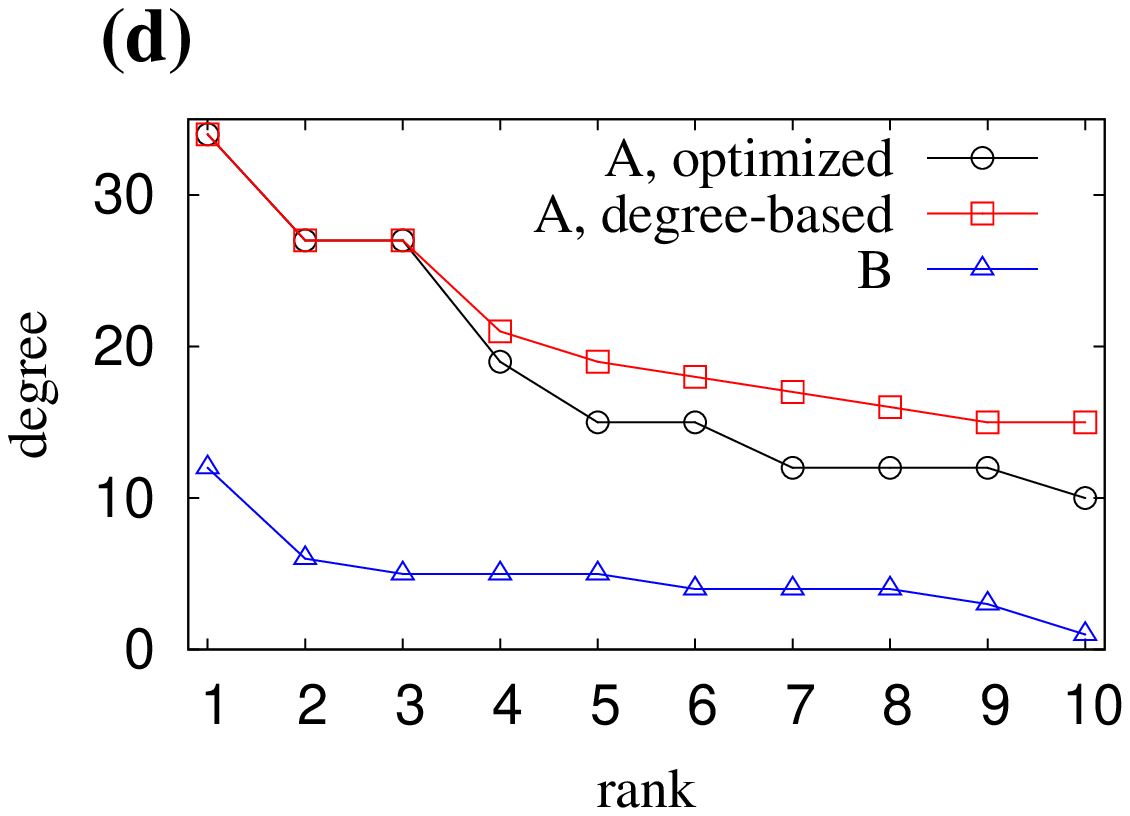}
\caption{Results for the coauthorship network in network science with $N=379$ nodes and 914 links.
The results when each zealot controls a single node are shown in (a) and (b). The control gain for both zealots is set to 1 in (a) and 10 in (b). 
The degrees of the controlled nodes when each zealot controls ten nodes are shown in (c) and (d) when the gain is equal to 1 and 10, respectively. See the caption of Fig.~\ref{fig:ba} for legends.}
\label{fig:netscience}
\end{center}
\end{figure}

As a second example, we examine a coauthorship network of network scientists \cite{Newman2006PhysRevE_finding}. The network is undirected and unweighted and has 1589 nodes and 2742 links.
We use the largest connected component of this network containing $N=379$ nodes and 914 links. The numerical procedure is the same as that used in the case of the BA model.
The numerical results are shown in Fig.~\ref{fig:netscience} and Table~\ref{tab:optimized}.
The general tendency is the same as that for the results for the BA model. However, the degree of the controlled node affects the results less strongly than in the case of the BA model. In fact, controlling various single nodes, including those with small degrees, yields a larger share of opinion A than controlling the largest-degree hub (Fig.~\ref{fig:netscience}). The difference between the share of A attained by the optimization and that attained by the degree-based protocol is also larger for the coauthorship network
than for the BA model (Table~\ref{tab:optimized}).

\subsubsection{Email communication network}

\begin{figure}[h]
\begin{center}
\includegraphics[width=8cm]{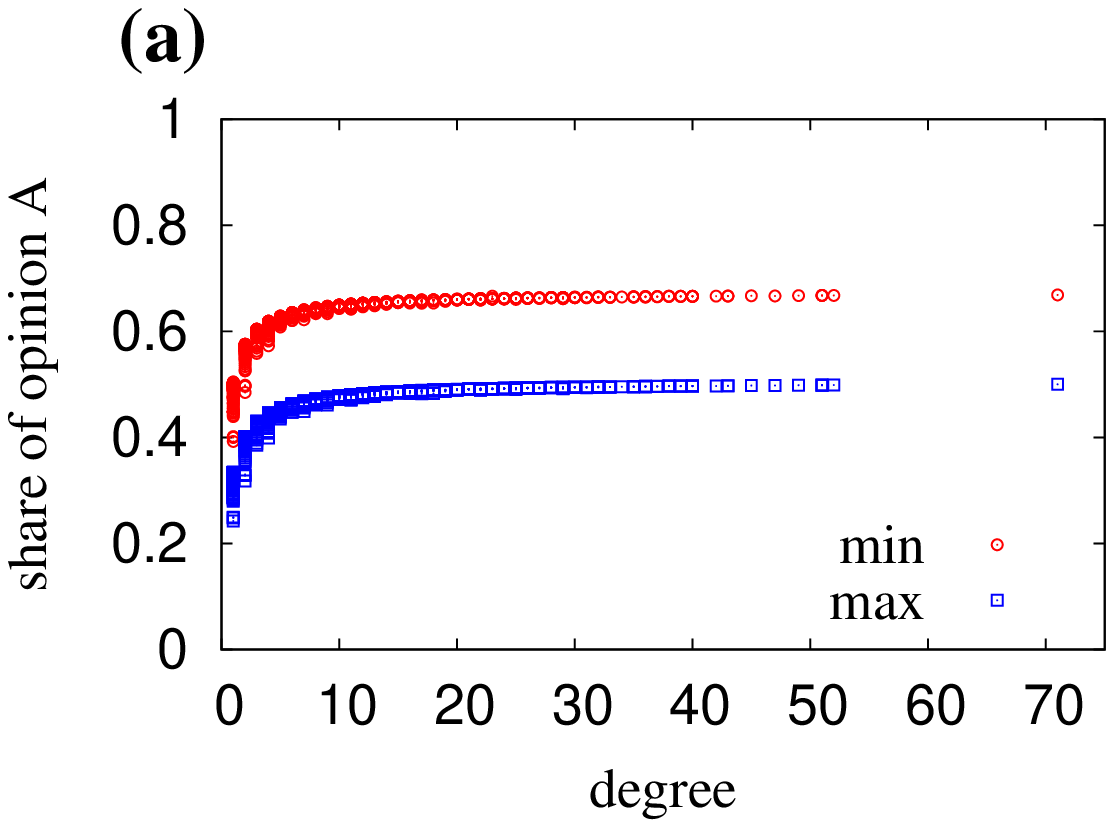}
\includegraphics[width=8cm]{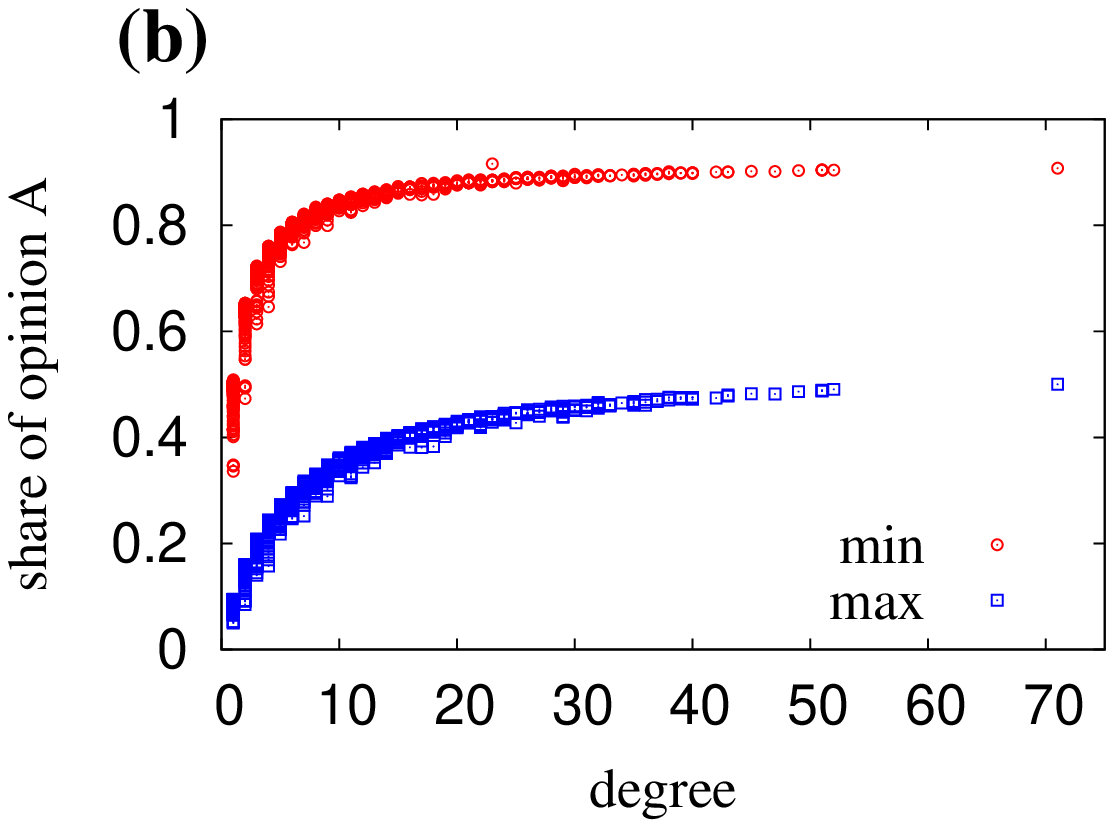}
\includegraphics[width=8cm]{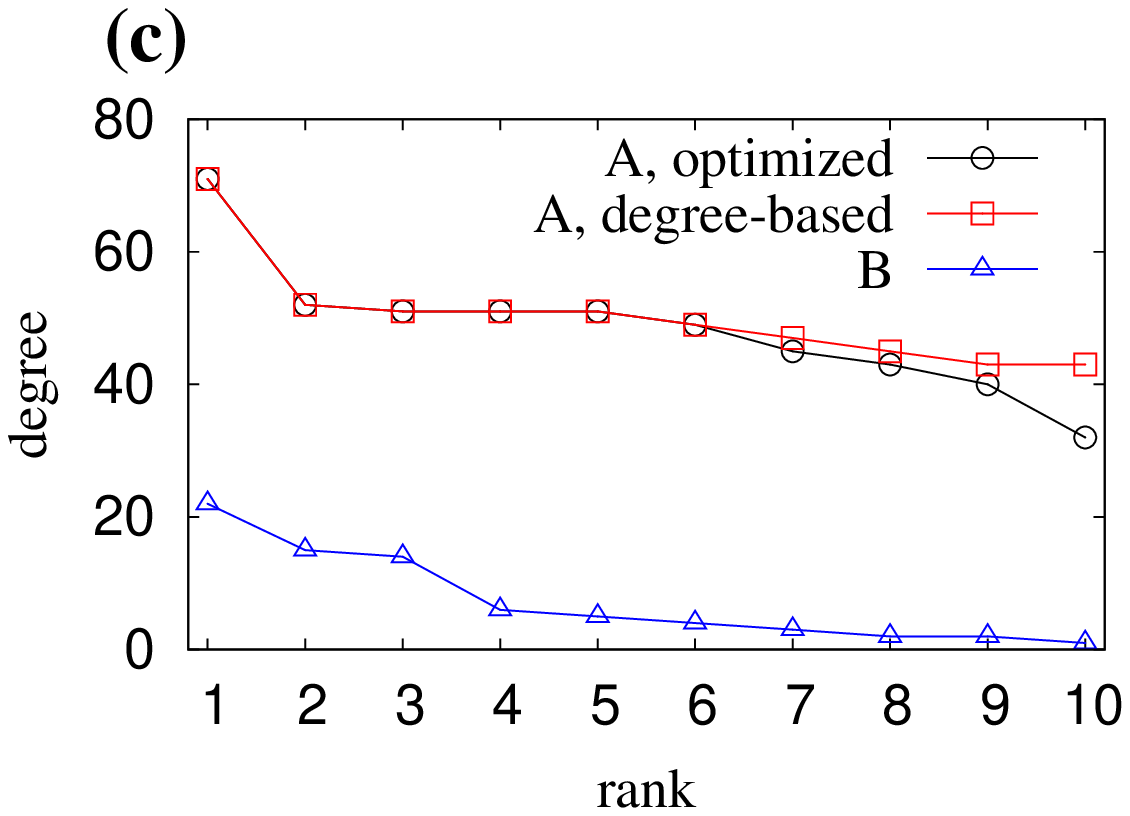}
\includegraphics[width=8cm]{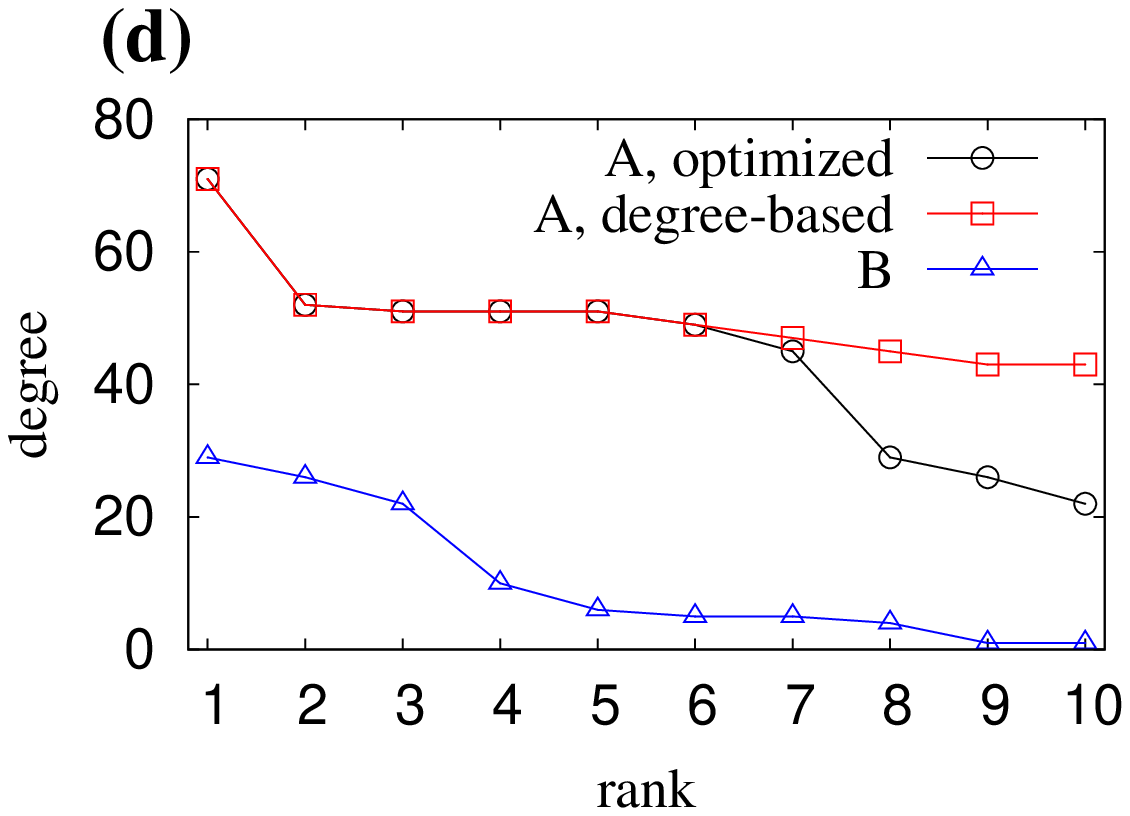}
\caption{Results for the email communication network with $N=1133$ nodes and 5451 links.
The results when each zealot controls a single node are shown in (a) and (b). 
The degrees of the controlled nodes when each zealot controls ten nodes are shown in (c) and (d).
In (a) and (c), the gain is set to 1. In (b) and (d), the gain is set to 10.
See the caption of Fig.~\ref{fig:ba} for legends.}
\label{fig:arenas}
\end{center}
\end{figure}

As a third example, we examine the largest connected component of email exchange network between members of the University Rovira i Virgili, Tarragona, Spain
\cite{GuimeraDanon2003PhysRevE}. This network is undirected and unweighted and has $N=1133$ nodes and 5451 links. The numerical results for this network are shown in Fig.~\ref{fig:arenas}
and Table~\ref{tab:optimized}. The correlation between the degree of the controlled node and the share of opinion A is very large for this network (Fig.~\ref{fig:arenas}). As a result, the share of A obtained after the optimization and that
based on the degree-based protocol are only slightly different (Table~\ref{tab:optimized}).

\subsubsection{Directed online social network}

All networks examine so far are undirected networks. In the voter model without zealots, the impact of node $i$ as measured by the fixation probability, i.e., the probability that the opinion starting solely from node $i$ is eventually adopted by all nodes in the network, is equal to the degree of node $i$ \cite{Antal2006PhysRevLett,Sood2008PhysRevE}. This may be the reason why the degree-based protocol is close to optimal for some undirected networks, i.e., the BA model and the email communication network. In contrast, the fixation probability for a node can substantially deviate from the out-degree (i.e., the number of other nodes that the focal node directly influences) in directed networks
\cite{MasudaOhtsuki2009NewJPhys}. Therefore, we postulate that the optimal nodes to be controlled may considerably deviate from high out-degree nodes in directed networks.

\begin{figure}[h]
\begin{center}
\includegraphics[width=8cm]{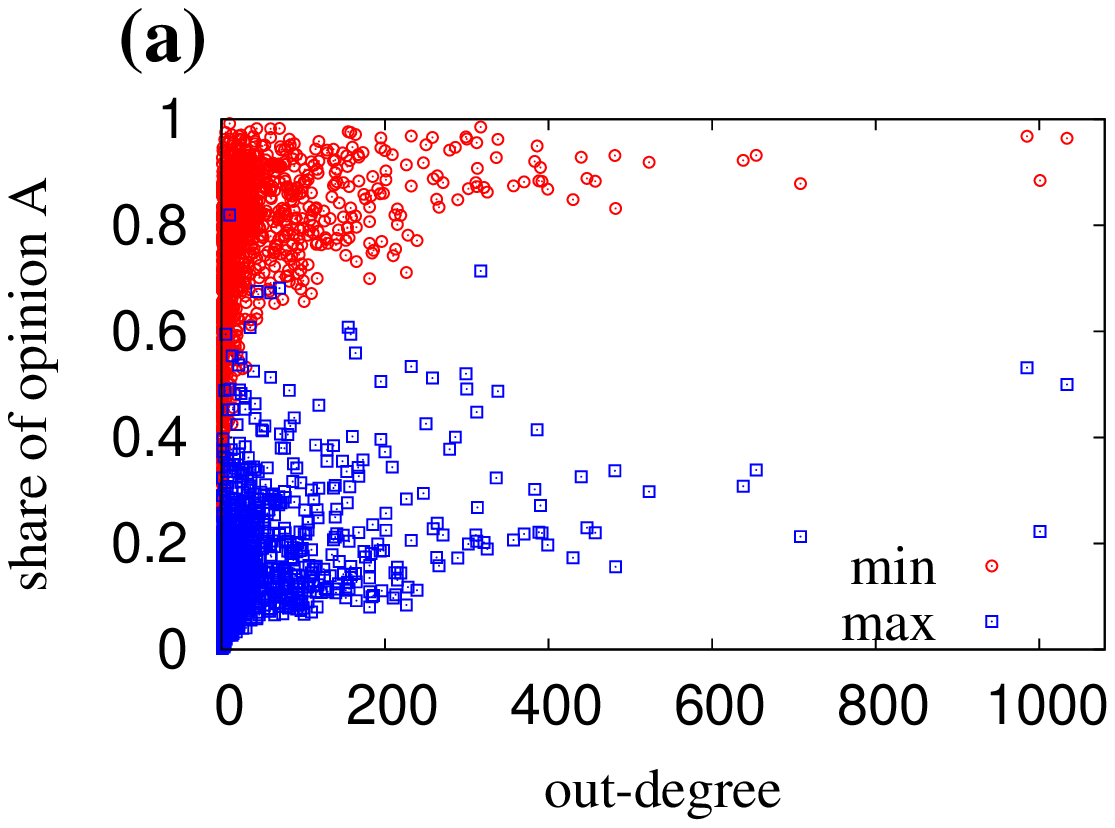}
\includegraphics[width=8cm]{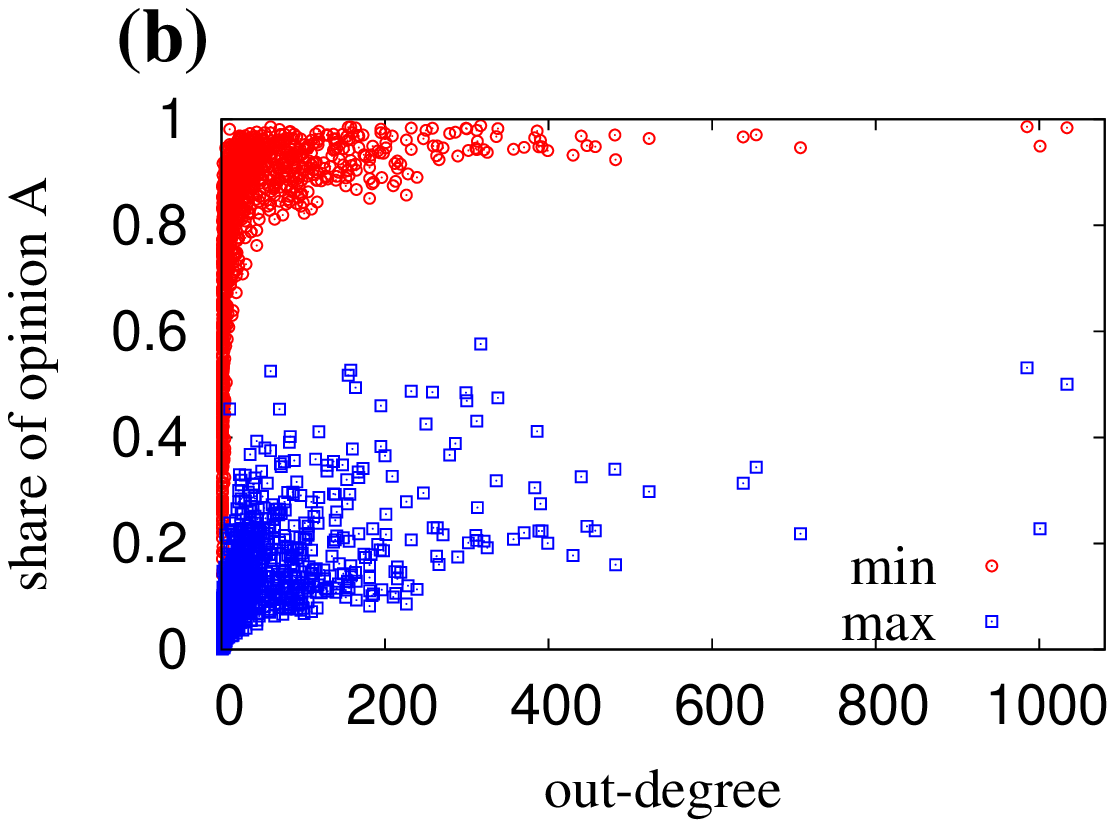}
\includegraphics[width=8cm]{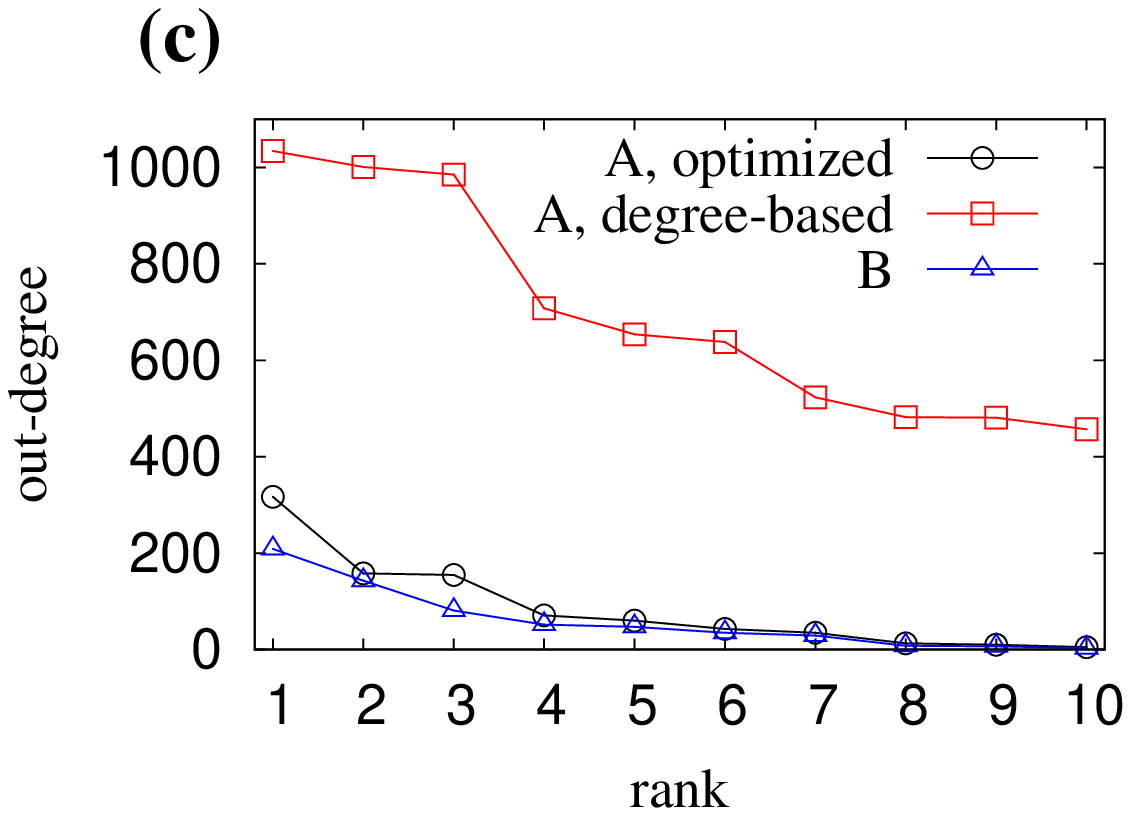}
\includegraphics[width=8cm]{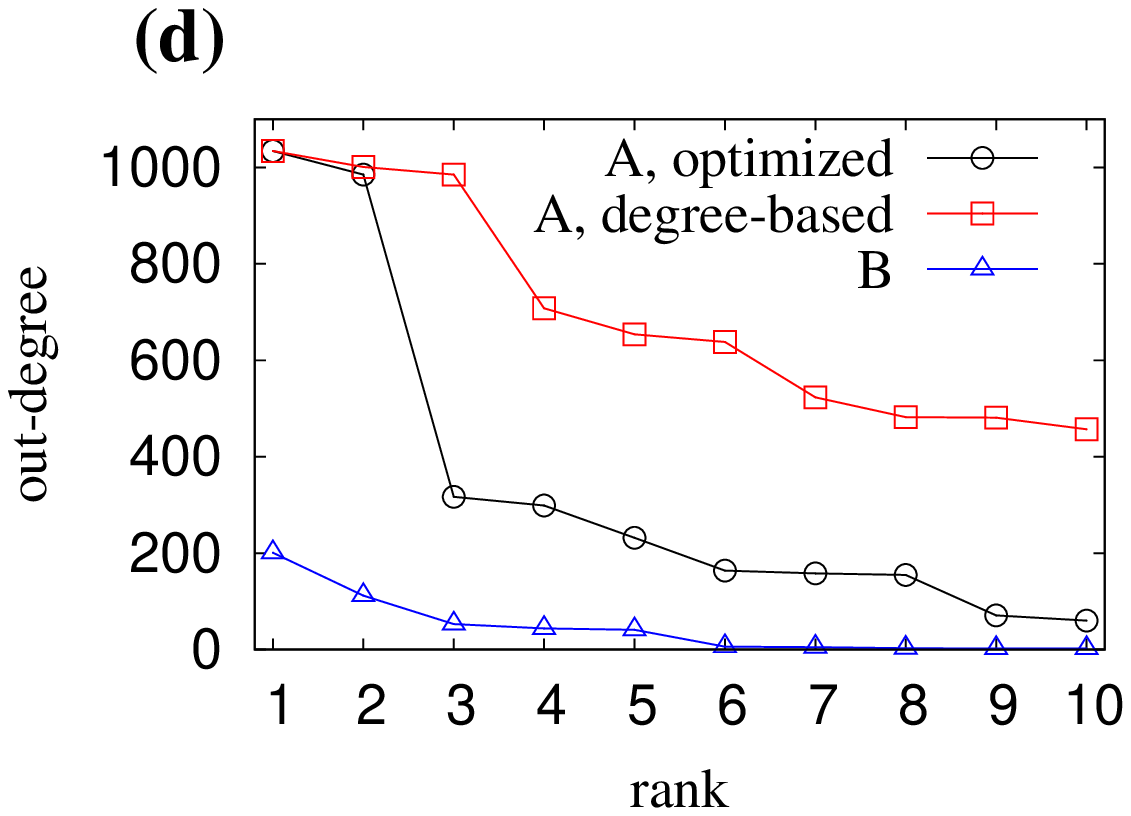}
\caption{Results for the online social network with $N=1294$ nodes and 19026 links.
The results when each zealot controls a single node are shown in (a) and (b).
The out-degrees of the controlled nodes when each zealot controls ten nodes are shown in (c) and (d).
In (a) and (c), the gain is set to 1. In (b) and (d), the gain is set to 10.
See the caption of Fig.~\ref{fig:ba} for legends.}
\label{fig:OC}
\end{center}
\end{figure}

To examine this possibility, we use an online social network  
among students at University of California, Irvine \cite{Panzarasa09JASIST}.
The network has $N=1899$ nodes and 20296 directed and weighted links.
We focus on the largest strongly connected component of this network containing
1294 nodes and 19026 weighted links. The numerical results are shown in Fig.~\ref{fig:OC} and Table~\ref{tab:optimized}.
We increased the number of iterations in the optimization procedure to $5\times 10^4$ because the $S_A$ value
was still increasing to some extent at the $2\times 10^4$th iterate.
The results indicate that the out-degree is not a good predictor of the impact of the node when it is controlled.

\section{Discussion}\label{sec:discussion}

We studied a problem of maximizing votes in opinion formation dynamics in complex networks when two opposing parties influence subsets of nodes. We proposed a heuristic algorithm based on exact counting of the mean vote and applied it to
artificial and real complex networks. We showed that the degree of the controlled node
was a main determinant of the efficiency of control in undirected networks, which is consistent with previous results \cite{Kuhlman2013CompNetw}. Controlling hubs is generally a good strategy. However, optimized selection of the controlled nodes realized a larger share of the desired opinion than the degree-based protocol did. The difference between the performance of the two methods was particularly large in directed networks.

We used the VM rule for opinion updating. In non-regular networks (i.e., networks in which some nodes have different degrees),
different update rules, namely, the invasion process (IP) and link dynamics (LD),
substantially change the outcome of opinion formation dynamics in the absence of zealots\cite{Castellano2005AIPConf,Antal2006PhysRevLett,Sood2008PhysRevE,MasudaOhtsuki2009NewJPhys}. Extending the present
framework to these cases is straightforward because the use of a different update rule corresponds to 
assigning a rescaled weight to each link \cite{MasudaOhtsuki2009NewJPhys}. In undirected networks without zealots, the fixation probability of the node is proportional to the degree under the VM rule, inversely proportional to the degree under the IP rule, and independent of the degree under the LD rule \cite{Antal2006PhysRevLett,Sood2008PhysRevE}. In the present study, we showed that, in the presence of zealots, the effect of the control is strongly correlated with the degree of the controlled node in undirected networks. Given the two results, it may be better to control small-degree nodes under the IP rule, and the degree of controlled nodes may be irrelevant under the LD rule.

There are several possible extensions of the present work. First, the proposed algorithm was not fast, in particular when the control gain was small. This was why our examples were relatively small networks (i.e., up to $N\approx 1300$ nodes). The bottleneck seems to be the Jacobi iteration. In addition, we employed a greedy method for optimization, which may be also inefficient.
Perturbation methods with which to assess the change in the eigenvector upon an external change (i.e., a slightly altered set of controlled nodes in our case) \cite{Restrepo2006PhysRevLett,Masuda2009NewJPhys_immu,Milanese2010PhysRevE,WatanabeMasuda2010PhysRevE,Lambiotte2014JComplexNetw} 
and genetic algorithms \cite{Hegselmann2014SSRN,Eymann2014preprint}
may be useful for designing better optimization algorithms.

Second, we assumed during optimization that the opponent zealot (i.e., B) did not change the set of controlled nodes. Only the focal zealot (i.e., A) was allowed to strategically behave to update the set of controlled nodes. This is an important limitation of the present study. In more realistic situations, the optimal selection of controlled nodes should depend on what the opponent does. Allowing both zealots to strategically behave may be an interesting question. An obtained equilibrium may be formulated as a game-theoretic equilibrium.

Third, we assumed that the complete knowledge of the network of independent voters was available to the controllers. In practical situations such as real voting and social mobilization, this assumption would be violated. Methods applicable when only partial knowledge of the network is available are desired.

Fourth, our analysis was concerned with the mean vote count in the steady state. In fact, the opinion of each agent fluctuates between the two opinions even in the equilibrium. Previous studies quantified such fluctuations for meanfield populations \cite{WuHuberman2004arxiv,Chinellato2007arxiv,Mobilia2007JStatMech} and mathematically proved the presence of fluctuations under some conditions for networks \cite{AcemogluComo2011IEEEConf,Yildiz2013ACMEconComput}. Quantifying the fluctuations for general networks may be possible. In addition, actual elections may occur before the equilibrium is reached. Therefore, analysis of transient dynamics seems to be of practical importance. Related to this issue,
the convergence rate of opinion dynamics depending on the location of zealots was examined in recent literature \cite{Pirani2014AmerCtrlConf}.

Fifth, we considered the case of two opinions for simplicity. Extension to the case of more than two opinions is straightforward. A theory of node importance in directed networks when there are a general number of root node \cite{MasudaKori2010PhysRevE} may prove useful to this end.

Sixth, we did not explore effects of self loops although our formulation allowed them. An increase in the weight of a self loop makes the corresponding independent voter relatively deaf to others' opinions when updating its own opinion. With a large weight of a self loop, the independent voter plays a role close to that of a zealot.

Seventh, controlling a single node with a large gain (e.g., 10) and controlling many nodes with a small gain (e.g., 10 nodes with a unity gain) may make a difference. We did not explore this point. It should be noted that, even if the total gain exerted by the controller is the same, controlling many nodes with small gains may be practically more costly than controlling one node with a large gain \cite{Delellis2010IEEEMag}.

Last, effects of zealots have also been examined for other collective dynamics such as
a local majority vote model \cite{GalamJacobs2007PhysicaA},
the naming game 
\cite{XieSreenivasan2011PhysRevE,XieEmenheiser2012PlosOne,Verma2014PhysicaA}, Axelrod's model for cultural dissemination \cite{SinghSreenivasan2012PhysRevE}, a model with agents in a neutral position \cite{Marvel2012PhysRevLett}, and
the prisoner's dilemma game \cite{Masuda2012SciRep,Mobilia2012PhysRevE,NakajimaMasuda2015JMathBiol}.
Maximization of the vote by pinning control may be also relevant in these contexts.

\section*{Acknowledgments}

N.M. acknowledges the support provided through JST, CREST, and JST, ERATO, Kawarabayashi Large Graph Project.

\section*{Appendix A: Proof of $u_{A,i} + u_{B,i}=1$}

By summing Eq.~\eqref{eq:u_{A,i}} and its equivalent for opinion B, we obtain
\begin{equation}
\left[L + {\rm diag}\left(p_{A,1}+p_{B,1}, \ldots, p_{A,N}+p_{B,N}\right)\right]
\begin{pmatrix}u_{A,1}+u_{B,1}\\ \vdots\\ u_{A,N}+u_{B,N}\end{pmatrix}
=
\begin{pmatrix}p_{A,1}+p_{B,1}\\ \vdots\\ p_{A,N}+p_{B,N}\end{pmatrix}.
\label{eq:u_{A,i}+u_{B,i}}
\end{equation}
Because $L + {\rm diag}\left(p_{A,1}+p_{B,1}, \ldots, p_{A,N}+p_{B,N}\right)$ is diagonally dominant,
Eq.~\eqref{eq:u_{A,i}+u_{B,i}} has a unique solution for unknowns
$u_{A,1}+u_{B,1}$, $\ldots$, $u_{A,N}+u_{B,N}$. In fact,
$u_{A,i}+u_{B,i}=1$ ($1\le i\le N$) solves Eq.~\eqref{eq:u_{A,i}+u_{B,i}} because $L$ is a Laplacian matrix
such that it has a zero right eigenvector $(1, \ldots, 1)^{\top}$, where $\top$ denotes the transposition.
Therefore, $u_{A,i}+u_{B,i}=1$ ($1\le i\le N$) holds true.

\newpage


\newpage
\clearpage

\newpage
\clearpage

\newpage
\clearpage

\newpage
\clearpage

\newpage
\clearpage

\newpage
\clearpage

\newpage
\clearpage

\newpage
\clearpage

\begin{table}[h]
\caption{Share of opinion A when the A zealot has optimized the controlled nodes and when it has used the degree-based protocol. Each zealot controls ten nodes. BA: Barab\'{a}si-Albert model, netsci: coauthorship network of network scientists, email: email communication network, online: online social network.
We have used different sets of randomly selected ten nodes for the B zealot for different gain values.}
\label{tab:optimized}
\begin{center}
	\begin{tabular}{|c|c| x{1.8cm} x{1.8cm}|}\hline
		network & gain & optimized & degree-based \\ \hline
		 & 1 & 0.576 & 0.574\\
		BA & 10 & 0.713 & 0.703\\
		 & 100 & 0.810 & 0.793\\ \hline
		 & 1 & 0.599 & 0.580\\
		netsci & 10 & 0.744 & 0.686\\
		 & 100 & 0.869 & 0.765 \\ \hline
		 & 1 & 0.570 & 0.570\\
		email & 10 & 0.679 & 0.670\\
		 & 100 & 0.860 & 0.840\\ \hline
		 & 1 & 0.896 & 0.678\\
		online & 10 & 0.898 & 0.811\\
		 & 100 & 0.947 & 0.929\\ \hline
	\end{tabular}
\end{center}
\end{table}

\end{document}